# Magnetism from Conductors, and Enhanced Non-Linear Phenomena

JB Pendry, AJ Holden, DJ Robbins, and WJ Stewart

*Abstract -* **We show that microstructures built from non-magnetic conducting sheets exhibit an effective magnetic permeability, $\mu_{eff}$, which can be tuned to values not accessible in naturally occurring materials, including large imaginary components of $\mu_{eff}$. The microstructure is on a scale much less than the wavelength of radiation, is not resolved by incident microwaves, and uses a very low density of metal so that structures can be extremely lightweight. Most of the structures are resonant due to internal capacitance and inductance, and resonant enhancement combined with compression of electrical energy into a very small volume greatly enhances the energy density at critical locations in the structure, easily by factors of a million and possibly by much more. Weakly non-linear materials placed at these critical locations will show greatly enhanced effects raising the possibility of manufacturing active structures whose properties can be switched at will between many states.**

JB Pendry is with The Blackett Laboratory, Imperial College, London, SW7 2BZ, UK.
AJ Holden, DJ Robbins, and WJ Stewart are with GEC-Marconi Materials Technology Ltd, Caswell, Towcester, Northamptonshire, NN12 8EQ, UK.

I. INTRODUCTION

In a sense every material is a composite, even if the individual ingredients consist of atoms and molecules. The original objective in defining a permittivity, $\varepsilon$, and permeability, $\mu$, was to present an homogeneous view of the electromagnetic properties of a medium. Therefore it is only a small step to replace the atoms of the original concept with structure on a larger scale. We shall consider periodic structures defined by a unit cell of characteristic dimensions $a$. The contents of the cell will define the effective response of the system as a whole.

Clearly there must be some restrictions on the dimensions of the cell. If we are concerned about the response of the system to electromagnetic radiation of frequency $\omega$ the conditions are easy to define:

$$a << \lambda = 2\pi c_0 \omega^{-1} \tag{1}$$

If this condition were not obeyed there would be the possibility that internal structure of the medium could diffract as well as refract radiation giving the game away immediately. Long wavelength radiation is too myopic to detect internal structure and in this limit an effective permittivity and permeability is a valid concept. In the next section we shall discuss how the microstructure can be related to $\varepsilon_{eff}, \mu_{eff}$.

In an earlier paper [1] we showed how a structure consisting of very thin infinitely long metal wires arranged in a 3D cubic lattice could model the response of a dilute plasma, giving a negative $\varepsilon_{eff}$ below a plasma frequency somewhere in the gigahertz range. Theoretical analysis of this structure has been confirmed by experiment [2]. Sievenpiper et al have also investigated plasma-like effects in metallic structures [3,4].

Ideally we should like to proceed in the magnetic case by finding the magnetic analogue of a good electrical conductor: unfortunately there isn't one! Nevertheless we can find some alternatives which we believe do give rise to interesting magnetic effects.

Why should we go to the trouble of microstructuring a material simply to generate a particular $\mu_{eff}$?

The answer is that atoms and molecules prove to be a rather restrictive set of elements from which to build a magnetic material. This is particularly true at frequencies in the gigahertz range where the magnetic response of most materials is beginning to tail off. Those materials, such as the ferrites, that remain moderately active are often heavy, and may not have very desirable mechanical properties. In contrast, we shall show, microstructured materials can be designed with considerable magnetic activity, both diamagnetic and paramagnetic, and can if desired be made extremely light.

There is another quite different motivation. We shall see that strong magnetic activity implies strongly inhomogeneous fields inside the material. In some instances this may result in local field strengths many orders of magnitude larger than in free space. Doping the composite with non linear material at the critical locations of field concentration gives enhanced non-linearity, reducing power requirements by the field enhancement factor. This is not an option available in a conventional magnetic material.

We show first how to calculate $\mu_{eff}$ for a system, then we propose some model structures which have magnetic activity and give some numbers for these systems. Finally we show how electrostatic energy can be strongly concentrated in these structures and hence the demonstrate potential for enhancing non linear effects.

## II. DEFINING AN EFFECTIVE PERMEABILITY

We are seeking to build structures with effective epsilon and mu,

$$\begin{aligned} \mathbf{B}_{ave} &= \mu_{eff}\mu_0 \mathbf{H}_{ave} \\ \mathbf{D}_{ave} &= \varepsilon_{eff}\varepsilon_0 \mathbf{E}_{ave} \end{aligned} \qquad (2)$$

where we assume that the structure is on a scale much shorter than the wavelength of any radiation so that we can sensibly speak of an average value for all the fields. A key question is how do the averages differ? Clearly if the structure is made of *thin* wires or sheets of metal then if the averages were taken over the same regions of space, $\varepsilon_{eff}, \mu_{eff}$ would always be unity. However we observe that Maxwell's equations,

$$\begin{aligned} \nabla \times \mathbf{H} &= +\partial \mathbf{D}/\partial t \\ \nabla \times \mathbf{E} &= -\partial \mathbf{B}/\partial t \end{aligned} \qquad (3)$$

may be applied in the integral form,

$$\begin{aligned} \int_c \mathbf{H} \cdot \mathbf{dl} &= +\frac{\partial}{\partial t}\int_s \mathbf{D}.\mathbf{dS} \\ \int_c \mathbf{E} \cdot \mathbf{dl} &= -\frac{\partial}{\partial t}\int_s \mathbf{B}.\mathbf{dS} \end{aligned} \qquad (4)$$

where the line integral is taken over a loop 'c' which encloses an area 's'.

This form of the equations immediately suggests a prescription for averaging the fields. For simplicity we shall assume that the periodic structure is described by a unit cell whose axes are orthogonal as shown in figure 1 below. Some of the arguments used in this section are similar to those we used in deriving a finite difference model of Maxwell's equations [5].

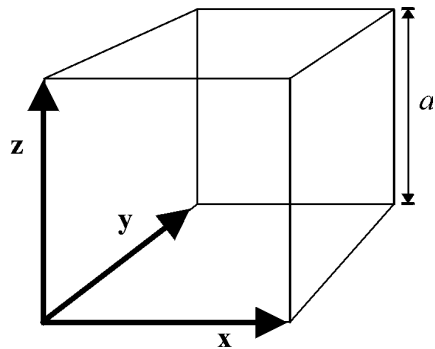

*Figure* 1. Unit cell of a periodic structure. We assume that the unit cell dimensions are much smaller that the wavelength of radiation, and average over local variations of the fields. In the case of the **B** - field we average over the faces of the cell and in the case of the **H** - field, over one of the edges.

We choose to define the components of $\mathbf{H}_{ave}$ by averaging the $\mathbf{H}-$field along each of the three axes of the unit cell. If we assume a simple cubic system,

$$\left(H_{ave}\right)_x = a^{-1} \int_{\mathbf{r}=(0,0,0)}^{\mathbf{r}=(a,0,0)} \mathbf{H} \cdot \mathbf{dr}$$

$$\left(H_{ave}\right)_y = a^{-1} \int_{\mathbf{r}=(0,0,0)}^{\mathbf{r}=(0,a,0)} \mathbf{H} \cdot \mathbf{dr} \qquad (5)$$

$$\left(H_{ave}\right)_z = a^{-1} \int_{\mathbf{r}=(0,0,0)}^{\mathbf{r}=(0,0,a)} \mathbf{H} \cdot \mathbf{dr}$$

There is only one caveat concerning the definition of the unit cell: its edges must not intersect with any of the structures contained within the unit cell. This leaves us free to cut the structure into a whole number of unit cells when we come to create a surface and ensures that the parallel component of $\mathbf{H}_{ave}$ is continuous across the surface as required in a consistent theory of an effective medium.

To define $\mathbf{B}_{ave}$ we average the $\mathbf{B}$–field over each of the three faces of the unit cell defined as follows:

$S_x$ is the surface defined by the vectors $\mathbf{y}, \mathbf{z}$
$S_y$ is the surface defined by the vectors $\mathbf{x}, \mathbf{z}$
$S_z$ is the surface defined by the vectors $\mathbf{x}, \mathbf{y}$

Hence we define,

$$\left(B_{ave}\right)_x = a^{-2} \int_{S_x} \mathbf{B} \cdot \mathbf{dS}$$

$$\left(B_{ave}\right)_y = a^{-2} \int_{S_y} \mathbf{B} \cdot \mathbf{dS} \qquad (6)$$

$$\left(B_{ave}\right)_z = a^{-2} \int_{S_z} \mathbf{B} \cdot \mathbf{dS}$$

The ratio defines the effective epsilon and mu from (2),

$$\left(\mu_{eff}\right)_x = \left(B_{ave}\right)_x / \left(\mu_0 H_{ave}\right)_x$$

$$\left(\mu_{eff}\right)_y = \left(B_{ave}\right)_y / \left(\mu_0 H_{ave}\right)_y \qquad (7)$$

$$\left(\mu_{eff}\right)_z = \left(B_{ave}\right)_z / \left(\mu_0 H_{ave}\right)_z$$

Thus if we seek a large effect we must try to create fields that are as inhomogeneous as possible.

We shall explore various configurations of thin sheets of metal, derive $\mu_{eff}$, and discuss the results with a view to making the effect as large as possible.

## III. EXAMPLES OF MAGNETIC MICROSTRUCTURES

### A. An Array of Cylinders

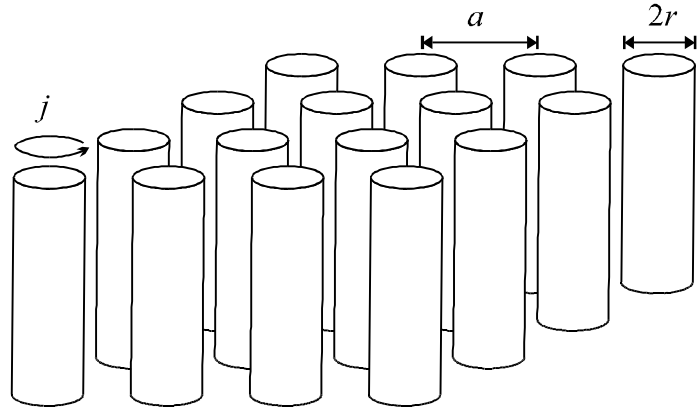

*Figure* 2. Model *A* consists of a square array of metallic cylinders designed to have magnetic properties in the direction parallel to the axes of the cylinders.

We start with a very simple structure for the purposes of illustration. Let us apply an external field, $H_0$, which we shall take to be parallel to the cylinders. We assume that the cylinders have a conducting surface so that a current, $j$, per unit length, flows. The field inside the cylinders is,

$$H = H_0 + j - \frac{\pi r^2}{a^2} j \tag{8}$$

where the second term on the right hand side is the field caused directly by the current, and the third term is the result of the depolarising fields with sources at the remote ends of the cylinders. If the cylinders are very long the depolarising field will be uniformly spread over the unit cell, but will have the same number of lines of force in it as the direct field inside the cylinders. We now calculate the total *emf* around the circumference of a cylinder:

$$\begin{aligned} emf &= -\pi r^2 \mu_0 \frac{\partial}{\partial t}\left[ H_0 + j - \frac{\pi r^2}{a^2} j \right] - 2\pi r \sigma j \\ &= +i\omega\pi r^2 \mu_0 \left[ H_0 + j - \frac{\pi r^2}{a^2} j \right] - 2\pi r \sigma j \end{aligned} \tag{9}$$

where $\sigma$ is the resistance of the cylinder surface per unit area. The net *emf* must balance and therefore,

$$j = \frac{-i\omega\pi r^2 \mu_0 H_0}{i\omega\pi r^2 \mu_0 \left[1 - \frac{\pi r^2}{a^2}\right] - 2\pi r \sigma} = \frac{-H_0}{\left[1 - \frac{\pi r^2}{a^2}\right] + i\frac{2r\sigma}{\omega r^2 \mu_0}} \tag{10}$$

We are now in a position to calculate the relevant averages. The average of the *B*-field over the entire unit cell is,

$$B_{ave} = \mu_0 H_0 \tag{11}$$

However if we average the *H*-field over a line lying entirely outside the cylinders,

$$H_{ave} = H_0 - \frac{\pi r^2}{a^2} j = H_0 - \frac{\pi r^2}{a^2} \frac{-H_0}{\left[1 - \frac{\pi r^2}{a^2}\right] + i\frac{2r\sigma}{\omega r^2 \mu_0}} = H_0 \frac{1 + i\frac{2\sigma}{\omega r \mu_0}}{\left[1 - \frac{\pi r^2}{a^2}\right] + i\frac{2\sigma}{\omega r \mu_0}} \qquad (12)$$

Hence we define,

$$\mu_{eff} = \frac{B_{ave}}{\mu_0 H_{ave}} = \frac{1 - \frac{\pi r^2}{a^2} + i\frac{2\sigma}{\omega r \mu_0}}{1 + i\frac{2\sigma}{\omega r \mu_0}} = 1 - \frac{\pi r^2}{a^2}\left[1 + i\frac{2\sigma}{\omega r \mu_0}\right]^{-1} \qquad (13)$$

For an infinitely conducting cylinder, or in the high frequency limit, $\mu_{eff}$ is reduced by the ratio of the cylinder volume to the cell volume. This ratio of volumes will turn out to be the key factor in determining the strength of the effect in all our models. Evidently in the present model $\mu_{eff}$ can never be less than zero, or greater than unity. It should also be mentioned that to maximise the effect we could have replaced the metallic cylinders with prisms of square cross section to maximise the volume enclosed within the prism.

If the resistivity of the sheets is high then the additional contribution to $\mu_{eff}$ is imaginary but always less than unity,

$$\mu_{eff} \approx 1 + i\frac{\pi r^3 \omega \mu_0}{2\sigma a^2}, \quad \sigma \gg \omega r \mu_0 \qquad (14)$$

A further point that should be noted is that all the structures we discuss have electrical as well as magnetic properties. In this particular case we can crudely estimate for electric fields perpendicular to the cylinders,

$$\varepsilon_{eff} = (1-F)^{-1} = \left(1 - \frac{\pi r^2}{a^2}\right)^{-1} \qquad (15)$$

where $F$ is the fraction of the structure not internal to a cylinder. In deriving (15) we assume that the cylinder is a perfect conductor, and neglect depolarising fields arising from interaction between cylinders. Inclusion of $\varepsilon_{eff}$ in our calculations removes one difficulty by ensuring that,

$$\lim_{\omega \to \infty} c_{light} = \lim_{\omega \to \infty} c_0 (\varepsilon_{eff} \mu_{eff})^{-\frac{1}{2}} = c_0 \qquad (16)$$

Evidently without $\varepsilon_{eff}$ the velocity of light in the effective medium would have exceeded that in free space. Most of the structures discussed in this paper have a similar $\varepsilon_{eff}$.

## B. A Capacitative Array of Sheets Wound on Cylinders

The previous structure showed a limited magnetic effect. Now we show how to extend the range of magnetic properties available to us by introducing capacitative elements into the structure. We take the same structure of cylinders as before except that the cylinders are now built in a 'split ring' configuration shown below in figure 3.

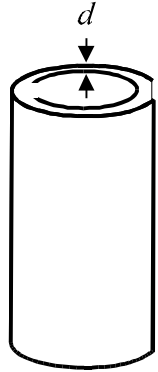

*Figure* 3. Model *B* consists of a square array of cylinders as for model *A* but with the difference that the cylinders now have internal structure The sheets are divided into a 'split ring' structure and separated from each other by a distance $d$. In any one sheet there is a gap which prevents current from flowing around that ring.

The important point is that there is a gap which prevents current from flowing around any one ring. However there is a considerable capacitance between the two rings which enables current to flow,

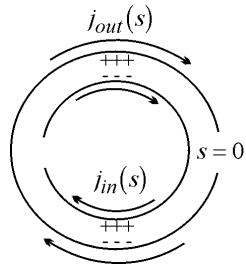

*Figure* 4. When a magnetic field parallel to the cylinder is switched on it induces currents in the 'split rings' as shown in the figure. The greater the capacitance between the sheets, the greater the current.

Detailed calculations give,

$$\mu_{eff} = 1 - \frac{F}{1 + \frac{2\sigma i}{\omega r \mu_0} - \frac{3}{\pi^2 \mu_0 \omega^2 C r^3}} \tag{17}$$

where $F$ is the fractional volume of the cell side occupied by the interior of the cylinder,

$$F = \frac{\pi r^2}{a^2} \tag{18}$$

and $C$ is the capacitance per unit area between the two sheets,

$$C = \frac{\varepsilon_0}{d} = \frac{1}{dc_0^2 \mu_0} \tag{19}$$

Hence,

$$\mu_{eff} = 1 - \frac{\frac{\pi r^2}{a^2}}{1 + \frac{2\sigma i}{\omega r \mu_0} - \frac{3dc_0^2}{\pi^2 \omega^2 r^3}} \tag{20}$$

Because we now have capacitance in the system which can balance the inductance present, $\mu_{eff}$ has a resonant form which we sketch below in figure 5.

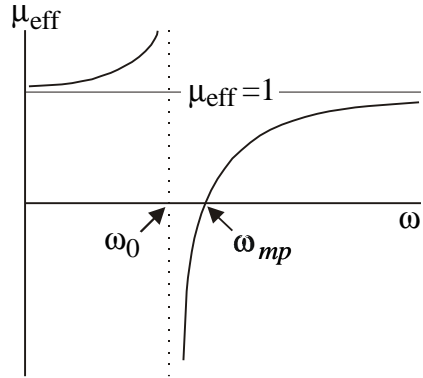

*Figure* 5. The effective magnetic permeability for *model B* shows a resonant structure dictated by the capacitance between the sheets and the magnetic inductance of the cylinder. We sketch the typical form for a highly conducting sample, $\sigma \approx 0$. Below the resonant frequency $\mu_{eff}$ is enhanced, but above resonance $\mu_{eff}$ is less than unity and may be negative close to the resonance.

Figure 5 illustrates the generic form of $\mu_{eff}$ for all the structures we present here.

We define $\omega_0$ to be the frequency at which $\mu_{eff}$ diverges,

$$\omega_0 = \sqrt{\frac{3}{\pi^2 \mu_0 C r^3}} = \sqrt{\frac{3dc_0^2}{\pi^2 r^3}} \tag{21}$$

and $\omega_{mp}$ to be the 'magnetic plasma frequency'

$$\omega_{mp} = \sqrt{\frac{3}{\pi^2 \mu_0 C r^3 (1-F)}} = \sqrt{\frac{3dc_0^2}{\pi^2 r^3 \left(1 - \frac{\pi r^2}{a^2}\right)}} \tag{22}$$

Note that the separation between $\omega_0$ and $\omega_{mp}$, which is a measure of the range of frequencies over which we see a strong effect, is determined by

$$F = 1 - \frac{\pi r^2}{a^2} \tag{23}$$

the fraction of the structure not internal to a cylinder. As for the case *A*, the simple cylinder, the high frequency limit is given by,

$$\lim_{\omega \to \infty} \mu_{eff}(\omega) = 1 - \frac{\pi r^2}{a^2} \tag{24}$$

We mention in passing that the system sustains longitudinal magnetic modes at the magnetic plasma frequency, the analogue of the plasma modes of a gas of free electrical charges [6,7]. Of course, we have no free magnetic poles, only the appearance of such as currents around the cylinders make the cylinder ends appear to support free magnetic poles in the fashion of a bar magnet.

Together with $\varepsilon_{eff}$ given in equation (15), which is also applicable here, we can illustrate a generic dispersion relationship shown below in figure 6.

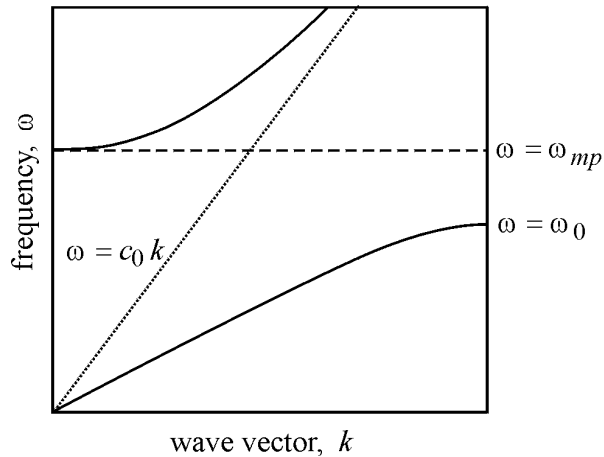

*Figure* 6. Generic dispersion relationship for resonant structures with a $\mu_{eff}$. The solid lines represent two-fold degenerate transverse modes, the dotted line a single longitudinal magnetic plasmon mode.

The relevant points to note are: (i) wherever $\mu_{eff}$ is negative there is a gap in the dispersion relationship, i.e. for,

$$\omega_0 < \omega < \omega_{mp} \tag{25}$$

(ii) a longitudinal magnetic plasma mode, dispersionless in this approximation, is seen at $\omega = \omega_{mp}$.
(iii) The dispersion relation converges asymptotically to the free space light cone. as discussed above. In fact metallic structures in general represent a fresh approach to the photonic insulator concept introduced independently by Yablonovitch [8,9] and John [10].

If we take the following values,

$$r = 2.0 \times 10^{-3} \, m$$
$$a = 5.0 \times 10^{-3} \, m \tag{26}$$
$$d = 1.0 \times 10^{-4} \, m$$

we get,

$$f_{mp} = (2\pi)^{-1} \omega_{mp} = 4.17 \times 10^9 \, \text{Hz} \tag{27}$$

$$f_0 = f_{mp} \sqrt{\left(1 - \frac{\pi r^2}{a^2}\right)} = 2.94 \times 10^9 \, \text{Hz} \tag{28}$$

Note that the frequency at which the structure is active corresponds to a free space wavelength of 10cm, much greater that the 0.5cm separation between cylinders. This will be typical of these capacitative structures and implies that the effective medium approximation will be excellent.

*C 'Swiss Roll' Capacitor*

We take the same arrangement of cylinders on a square lattice as before except that the cylinders are now build as follows:

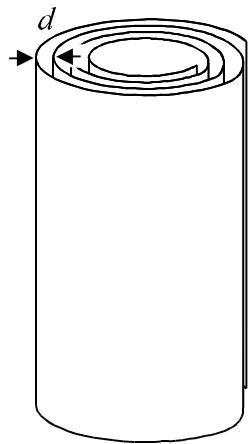

*Figure* 7. In model *C* a metallic sheet is wound around each cylinder in a coil. Each turn of the coil is spaced by a distance *d* from the previous sheet.

The important point is again that no current can flow around the coil except by virtue of the self capacitance,

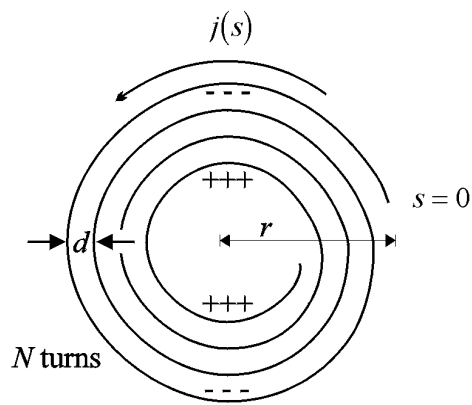

*Figure* 8. When a magnetic field parallel to the cylinder is switched on it induces currents in the coiled sheets as shown in the figure. Capacitance between the first and last turns of the coil enables current to flow.

In this instance we find for the effective permeability,

$$m_{eff} = 1 - \frac{F}{1 + \frac{2si}{wrm_0(N-1)} - \frac{1}{2p^2 r^3 m_0(N-1)^2 w^2 C}}$$

$$= 1 - \frac{\frac{p r^2}{a^2}}{1 + \frac{2si}{wrm_0(N-1)} - \frac{dc_0^2}{2p^2 r^3(N-1)w^2}}$$

(29)

where $F$ is as before the fraction of the structure not internal to a cylinder, and the capacitance per unit area between the first and the last of the coils is,

$$C = \frac{e_0}{d(N-1)} = \frac{1}{m_0 dc_0^2(N-1)}$$

(30)

The critical frequencies are,

$$w_0 = \sqrt{\frac{1}{2p^2 r^3 m_0 C(N-1)^2}} = \sqrt{\frac{dc_0^2}{2p^2 r^3(N-1)}}$$

(31)

$$w_{mp} = \sqrt{\frac{1}{F 2p^2 r^3 m_0 C(N-1)^2}} = \sqrt{\frac{dc_0^2}{\left(1 - \frac{p r^2}{a^2}\right) 2p^2 r^3(N-1)}}$$

(32)

If we take the values we used before in (26),

$$\begin{aligned} r &= 2.0 \times 10^{-3} \, \text{m} \\ a &= 5.0 \times 10^{-3} \, \text{m} \\ d &= 1.0 \times 10^{-4} \, \text{m} \\ N &= 11 \end{aligned}$$

(33)

we get,

$$f_0 = (2p)^{-1} w_0 = 0.380 \times 10^9 \, \text{Hz}$$

(34)

$$f_{mp} = (2p)^{-1} w_{mp} = 0.539 \times 10^9$$

(35)

i.e. there is much more capacitance in this model and the range of active frequencies is an order of magnitude lower than it was in model C which used only two overlapping sheets.

Choosing an even smaller scale and reducing the number of turns in order to drive up the frequencies to our range of interest,

$$r = 2.0 \times 10^{-4} \, m$$
$$a = 5.0 \times 10^{-4} \, m$$
$$d = 1.0 \times 10^{-5} \, m \qquad (36)$$
$$N = 3$$

we get,

$$f_0 = 8.50 \times 10^9 \, Hz \qquad (37)$$

$$f_{mp} = 12.05 \times 10^9 \, Hz \qquad (38)$$

Note that the free space wavelength at the plasma frequency is around 3cm and compare this to the very much smaller spacing between cylinders of 0.05cm.

We shall now calculate the dispersion of $\mu_{eff}$ for various parameters. First let us take the parameters given in equation (36). The resulting dispersion of $\mu_{eff}$ is shown below in figure 9

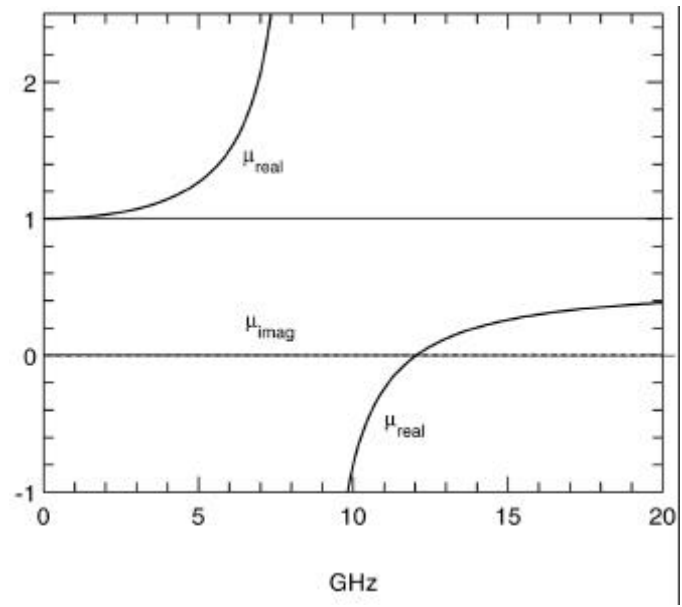

*Figure* 9. Dispersion with frequency of $\mu_{eff}$ for a Swiss roll structure, calculated for the parameters shown in equation (36), assuming that the metal has zero resistivity.

Next we enquire what is the effect of making the sheets resistive? Below we present a series of calculations for various values of the resistivity, $\sigma$, given in $\Omega$.

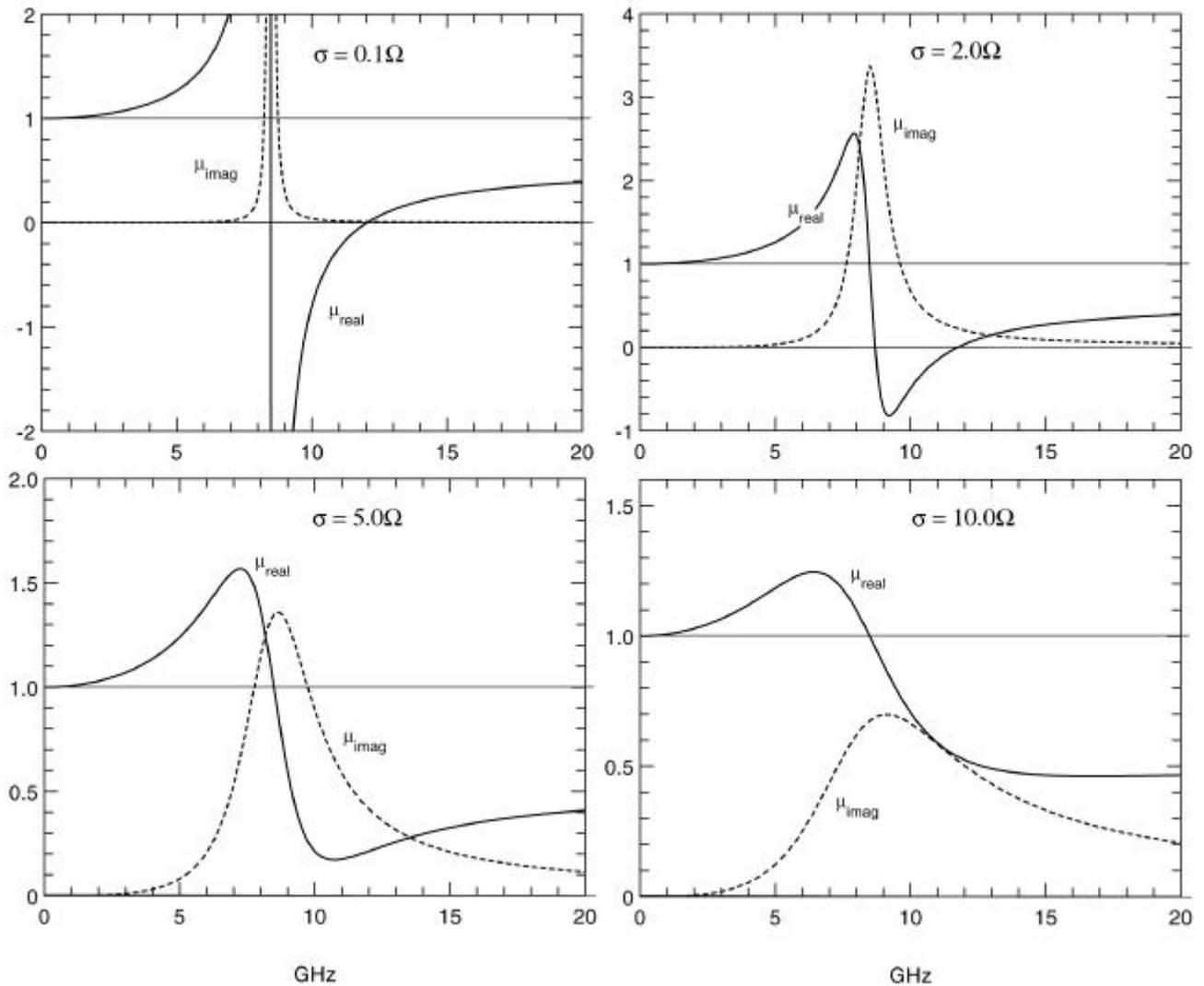

*Figure* 10. Dispersion with frequency of $\mu_{eff}$ for a Swiss roll structure, calculated for the parameters shown in equation (36), for various values of the resistivity of the sheets: $0.1\Omega$, $2.0\Omega$, $5.0\Omega$, $10.0\Omega$.

In figure 10 we increase the resistivity from $0.1\Omega$ to $10.0\Omega$. Note the broadening of the resonance, the complementary behaviour of $\mu_{real}$ and $\mu_{imag}$, dictated by Kramers Kronig, and how resistivity limits the maximum effect achieved.

Next we explore the dependence on the radius of the cylinders. In figure 11 the radius of the cylinders is decreased, reducing the volume fraction occupied by the cylinders, and raising the resonant frequency by a factor of two. We also decrease *d*, the spacing between the sheets, increasing the capacitance in the system and bringing the resonant frequency back down to its original value.

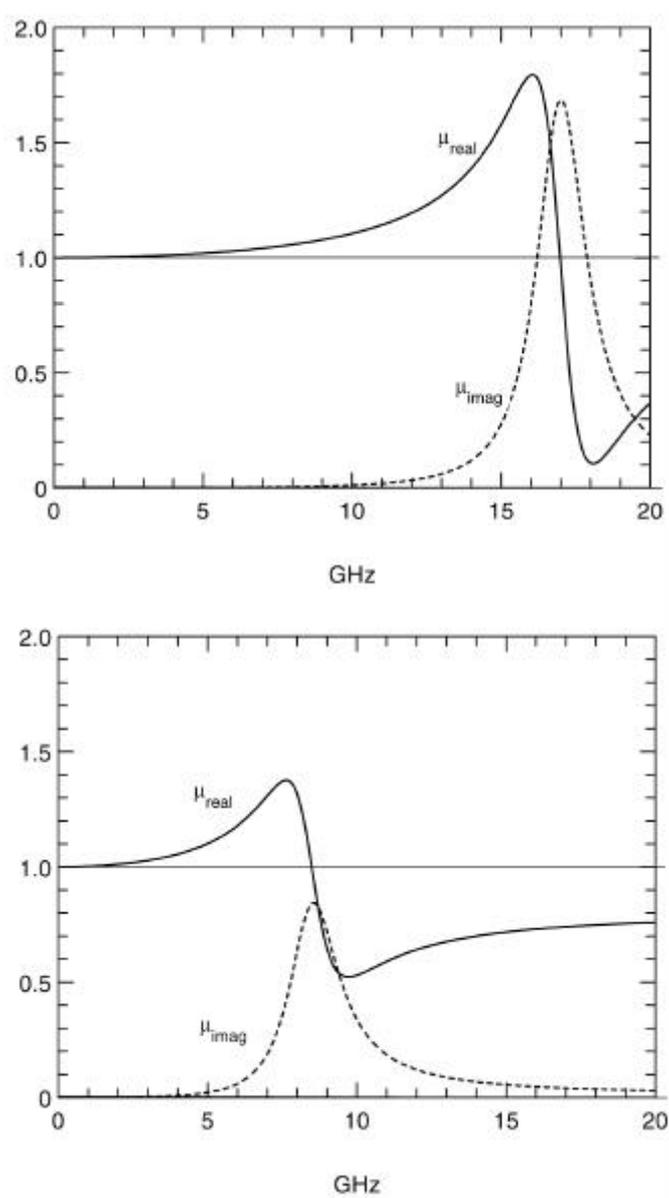

*Figure* 11. Dispersion with frequency of $\mu_{eff}$ for a Swiss roll structure. Top: calculated for the parameters shown in equation (36), except that the resistivity of the sheets is now $2.0\Omega$, and the radius of the cylinders has been reduced from $2.0 \times 10^{-4}\,\text{m}$ to $1.26 \times 10^{-4}\,\text{m}$, thus raising the resonant frequency by a factor of two. Bottom: $d$, the spacing between the sheets, has been reduced to $0.25 \times 10^{-5}\,\text{m}$ bringing the resonant frequency back to the original value.

Using capacitative cylindrical structures such as the Swiss roll structure we can adjust the magnetic permeability typically by a factor of two and, in addition if we desire, introduce an imaginary component of the order of unity. The latter implies that an electromagnetic wave moving in such a material would decay to half its intensity within a single wavelength. This presumes that we are seeking broad-band effects that persist over the greater part of the 2-20GHz region. However if we are prepared to settle for an effect over a narrow range of frequencies spectacular enhancements of the magnetic permeability can be achieved, limited only be the resistivity of the sheets and by how narrow a band we are willing to tolerate.

## III. AN ISOTROPIC MAGNETIC MATERIAL

The structures shown above give magnetic properties when the field is aligned along the axes of the cylinders, but have essentially zero magnetic response in other directions. They suffer from another potential problem: if the alternate polarisation is considered where the electric field is not parallel to the cylinders, the system responds like an effective metal because current is free to flow along the length of the cylinders. For some applications this highly anisotropic behaviour may be undesirable. Therefore we redesign the system with a view to restoring isotropy, and minimising purely electrical effects.

To this end we need a basic unit that is more easily packed into arrays than is a cylinder, and which avoids the continuous electrical path provided by a metal cylinder. We propose an adaptation of the 'split ring' structure in which the cylinder is replaced by a series of flat disks each of which retains the 'split ring' configuration but in slightly modified form: see figure 12. First we shall calculate the properties of disks stacked in a square array as shown if figure 13. This structure is still anisotropic, a problem we shall address in a moment, but by eliminating the continuous conducting path which the cylinders provided, it eliminates most of the electrical activity along this direction.

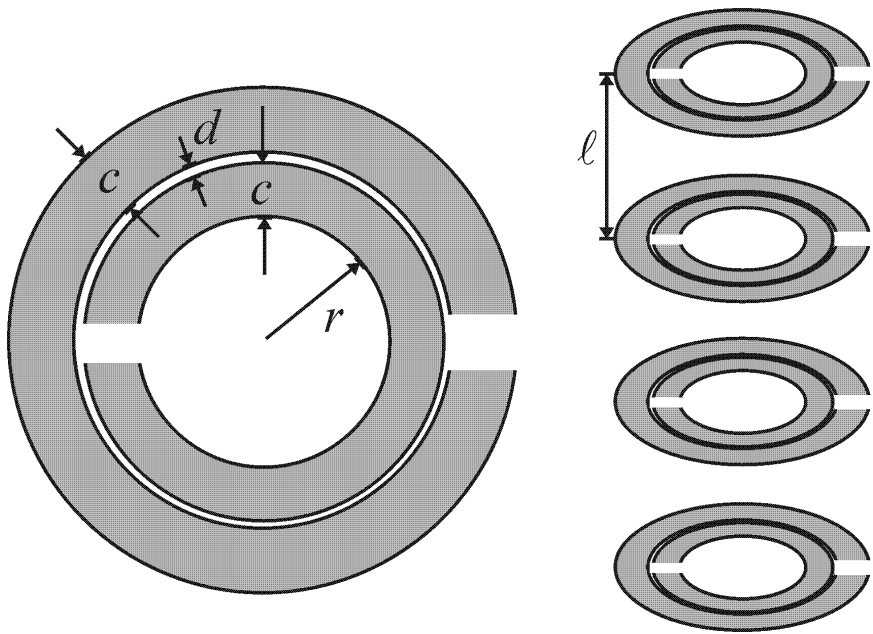

*Figure* 12. Left: a plan view of a split ring showing definitions of distances. Right a sequence of split rings shown in their stacking sequence. Each split ring comprises two thin sheets of metal. The ring shown is a scaled up version defined by the parameters shown below in figure 13.

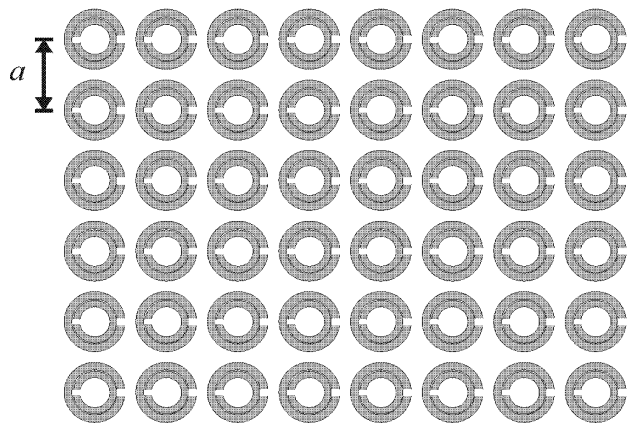

*Figure* 13. Plan view of a split ring structure in a square array, lattice spacing *a*.

The two dimensional square array shown in figure 13 can be made by printing with metallic inks. If each printed sheet is then fixed to a solid block of inert material, thickness $a$, the blocks can be stacked to give columns of rings. This would establish magnetic activity along the direction of stacking, the $z-$axis. The unit cell of this structure is shown in figure 14 on the left.

How do we make a symmetrical structure? Start from the structure just described comprising successive layers of rings stacked along the $z-$ axis. Next cut up the structure into a series of slabs thickness $a$, making incisions in the $y-z$ plane and being careful to avoid slicing through any of the rings. Each of the new slabs contains a layer of rings but now each ring is perpendicular to the plane of the slab and is embedded within. Print onto the surface of each slab another layer of rings and stack the slabs back together again. The unit cell of this second structure is shown in the middle of figure 14.

In the next step a third set of slabs is produced by cutting in the $x-z$ plane, printing on the surface of the slabs, and reassembling. Finally we now have a structure with cubic symmetry whose unit cell is shown on the right of figure 14.

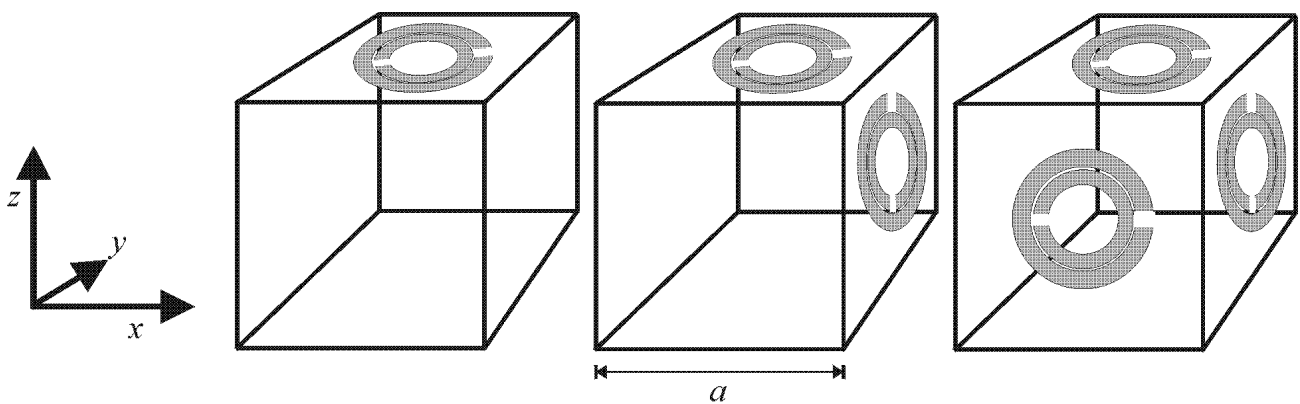

*Figure* 14. Building 3D symmetry: each successive re-stacking of the structure adds a ring to another side of the unit cell.

Of course an alternate method of manufacturing this structure would be to start from a set of cubes of the inert material and laboriously stick rings to their sides before assembling the cubes into a lattice. The cut and paste method we suggest above is much more efficient.

Now let us calculate the effective permeability. First we need to calculate the capacitance between the two elements of the split ring. We shall assume:

$$r \gg c, \quad r \gg d \tag{39}$$

$$\ell < r \tag{40}$$

$$\ln\frac{c}{d} \gg \pi \tag{41}$$

Under these conditions we can calculate the capacitance between unit length of two parallel sections of the metallic strips:

$$C_1 = \frac{\varepsilon_0}{\pi}\ln\frac{2c}{d} = \frac{1}{\pi\mu_0 c_0^2}\ln\frac{2c}{d} \tag{42}$$

The effective magnetic permeability we calculate on the assumption that the rings are sufficiently close together that the magnetic lines of force due to currents in the stacked rings, are essentially the same as those in a continuous cylinder. This can only be true if the radius of the rings is of the same order as the unit cell side. We arrive at:

$$\mu_{eff} = 1 - \frac{\frac{\pi r^2}{a^2}}{1 + \frac{2\ell\sigma_1}{\omega r\mu_0}i - \frac{3\ell}{\pi^2\mu_0\omega^2 C_1 r^3}} = 1 - \frac{\frac{\pi r^2}{a^2}}{1 + \frac{2\ell\sigma_1}{\omega r\mu_0}i - \frac{3\ell c_0^2}{\pi\omega^2\ln\frac{2c}{d} r^3}} \tag{43}$$

where $\sigma_1$ is the resistance of unit length of the sheets measured around the circumference.

To give some examples let us choose a convenient set of parameters:

$$\begin{aligned}
a &= 1.0 \times 10^{-2}\,\text{m} \\
c &= 1.0 \times 10^{-3}\,\text{m} \\
d &= 1.0 \times 10^{-4}\,\text{m} \\
\ell &= 2.0 \times 10^{-3}\,\text{m} \\
r &= 2.0 \times 10^{-3}\,\text{m}
\end{aligned} \tag{44}$$

Figures 12, 13 show the rings drawn to scale. These parameters do not quite satisfy all the inequalities, which is difficult to do with reasonable numbers, but note that the inequalities are only important to the accuracy of our formulae, not to the functioning of the structure. The resonant frequency at which $\mu_{eff}$ diverges is given by,

$$\omega_0^2 = \frac{3\ell c_0^2}{\pi\ln\frac{2c}{d} r^3} = 7.1 \times 10^{21} \tag{45}$$

or,

$$\omega = 2\pi \times 13.5 \text{GHz} \tag{46}$$

If we choose to manufacture the split rings from a layer of copper, it is easily possible to achieve $\sigma_1 \approx 200.0$. Evidently from figure 15, this produces a highly resonant structure.

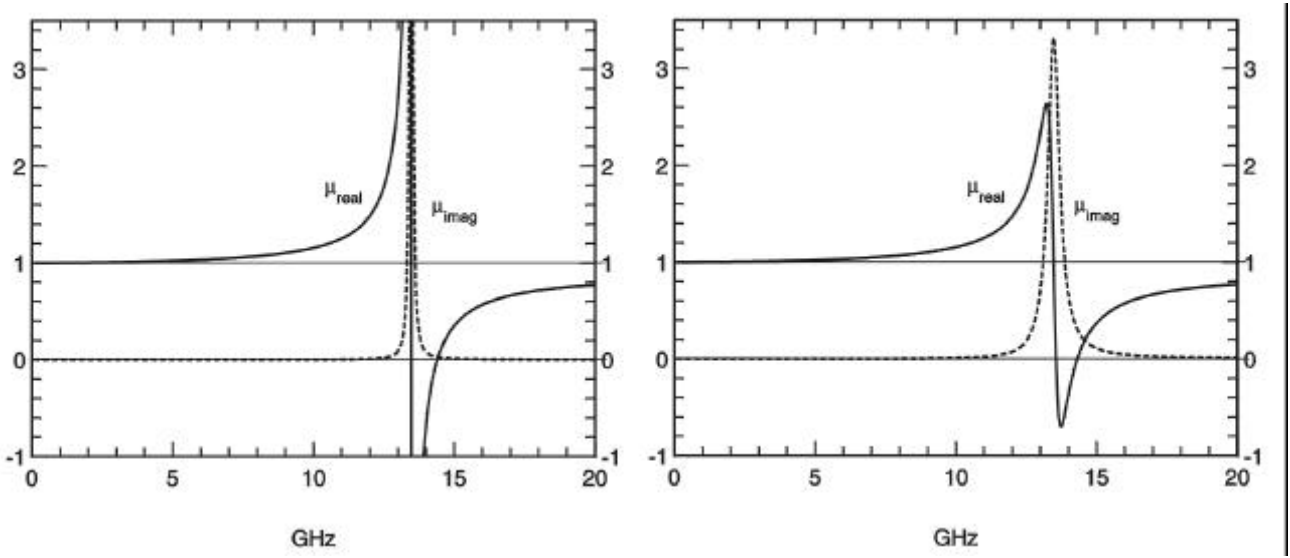

*Figure* 15. Plot of $\mu_{eff}$ for the cubic split ring structure calculated using the chosen parameters. Left: for copper rings, $\sigma_1 = 200.0$; right: for more resistive rings, $s_1 = 2000.0$.

In order to see a substantial effect we have to increase the resistance either by increasing the resistivity of the material of which the rings are made, or by making them thinner.

The scaling of frequency with size can be deduced from (45) we see that the resonant frequency scales uniformly with size: if we double the size of all elements in a given structure, the resonant frequency halves. Nearly all the critical properties are determined by this frequency.

IV. ENHANCED NON-LINEAR EFFECTS

We have seen how the addition of capacitance to the structure gives a far richer variety of magnetic behaviour. Typically this happens through a resonant interaction between the natural inductance of the structure and the capacitative elements, and at the resonant frequency electromagnetic energy is shared between the magnetic fields and the electrostatic fields within the capacitative structure. To put this more explicitly: take the split ring structure described in figures 12, 13: most of the electrostatic energy of the capacitor is located in the tiny gap between the rings. Concentrating most of the electromagnetic energy in this very small volume will results in an enormously enhanced energy density.

If we wish to enhance the non-linear behaviour of a given compound, we locate a small amount of the substance in the gap where the strong electrostatic fields are located. Since the response scales as the cube of the field amplitude, we can expect enhancements of the order of the energy density enhancement squared. Furthermore not only does the structure enhance the non-linearity, it does so in a manner that is very economical with the material: less that 1% of the structure need be filled with the non-linear substance.

Note that there is a symmetry between, on the one hand the present structures designed to generate a magnetic permeability and within which we find enhanced electrostatic fields, and on the other hand the earlier thin wire structures [1,2] designed to generate a negative electrical permittivity, and within which we find enhanced magnetic fields.

We shall now calculate the energy density in the capacitance between the two split rings in figures 12,13. First we calculate the voltage between the two rings as a function of the incident magnetic field, $H_0$.

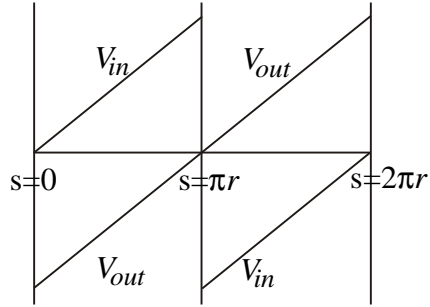

*Figure* 16. The *emf* acting around one of the sheets of the split ring in figure 12 as a function of the distance, $s$, around the ring. $V_{in}$ denotes the *emf* on the inner ring, and $V_{out}$ that on the outer ring. Note that this ring is cut at $s=0$ so that the *emf* is discontinuous.

The electric field between the two halves of the ring is then of the order,

$$E_{ring} \approx \pi V_0 d^{-1} \qquad (47)$$

We calculate that,

$$V_0 = \frac{3\ell i}{2wC_1\pi^2 r^2} \frac{-H_0}{\left[1-\frac{\pi^2}{a^2}\right] + \frac{2\ell \sigma_1 i}{wr\mu_0} - \frac{3\ell}{\pi^2 \mu_0 w^2 C_1 r^3}} \qquad (48)$$

Hence on substituting from (42) and (47) into (48):

$$E_{ring} \approx \frac{3\ell \mu_0 c_0^2 i}{2w\, dr \ln\frac{2c}{d}} \frac{-H_0}{\left[1-\frac{\pi^2}{a^2}\right] + \frac{2\ell \sigma_1 i}{wr\mu_0} - \frac{3\ell c_0^2}{\pi w^2 r^3 \ln\frac{2c}{d}}} \qquad (49)$$

Now we argue that the electrostatic energy density in the incident electromagnetic field is equal to the magnetic energy density, which in turn can be related to the electrostatic energy density in the ring. Hence,

$$\frac{\frac{1}{2}\varepsilon_0 |E_{ring}|^2}{\frac{1}{2}\mu_0 |H_{ave}|^2} = \frac{\varepsilon_0 \pi^2 r^2 d^{-2}}{\mu_0} \left| \frac{\frac{3\ell}{2\omega C_1 \pi^2 r^2}}{1 + \frac{2\ell \sigma_1 i}{\omega r \mu_0} - \frac{3\ell}{\pi^2 \mu_0 \omega^2 C_1 r^3}} \right|^2 \qquad (50)$$

If we evaluate this formula on resonance we get a much simplified formula,

$$\text{resonant enhancement} = Q = \frac{\frac{1}{2}\varepsilon_0 |E_{ring}(\omega_0)|^2}{\frac{1}{2}\mu_0 |H_0(\omega_0)|^2} = \left|\frac{\pi\omega^2 r^3 \mu_0}{4\ell\sigma d c_0}\right|^2 \quad (51)$$

Let us take as an example the parameters used to calculate figure 15,

$$\begin{aligned} d &= 1.0\times 10^{-4}\,\text{m} \\ \ell &= 2.0\times 10^{-3}\,\text{m} \\ r &= 2.0\times 10^{-3}\,\text{m} \\ \sigma_1 &= \frac{R}{ct} = 200.0 \\ \omega_0^2 &= 7.1\times 10^{21} \end{aligned} \quad (52)$$

Hence,

$$Q = \left[\frac{\pi\omega_0^2 r^3 \mu_0}{4\ell\sigma_1 d c_0}\right]^2 = 2.1853\times 10^7 \quad (53)$$

A more detail picture of enhancement as a function of frequency is shown in figure 17.

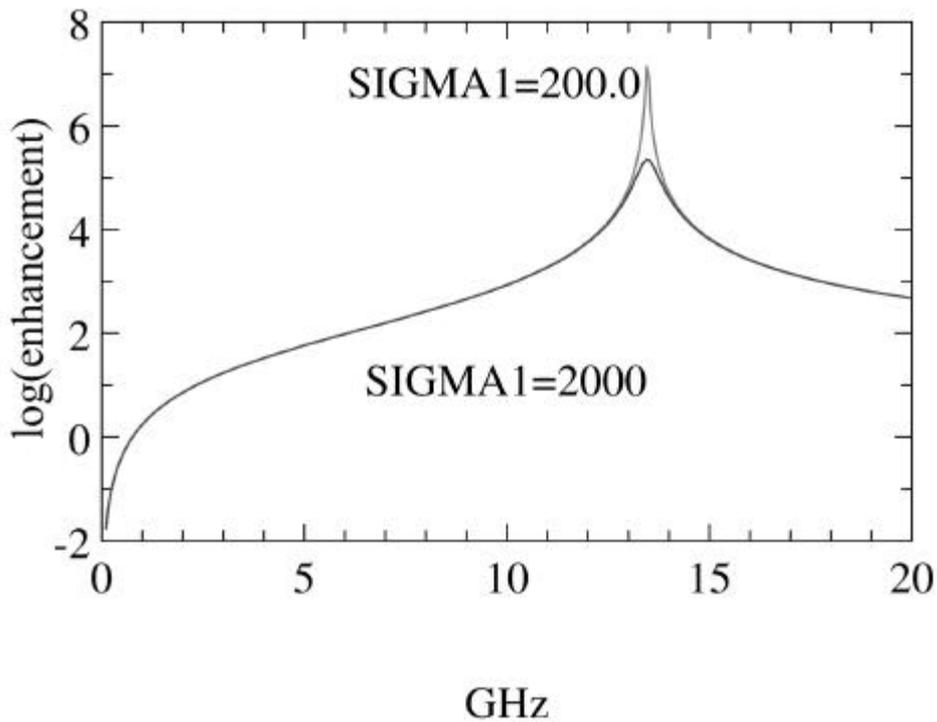

*Figure* 17. Enhancement of the energy density of the electric field within the gap between the split rings (see figures 12 and 13) for two different values of the resistivity of the metal sheet. The corresponding values of $\mu_{eff}$ are shown in figure 15.

For example: a beam of microwaves at 13.41GHz with power flux of $10^4\,\text{wm}^{-2}$ has an electric field strength of the order of $2\times 10^3\,\text{Vm}^{-1}$ in vacuo. If this beam were incident on, and entirely transmitted into, our magnetic structure it would generate a field strength of the order of $4 \times 10^{10} \, \text{Vm}^{-1}$ in the space between the split rings, or of the order of $10^6 \, \text{V}$ between the edges of the two rings: more than enough to cause electrical breakdown in air! It is evident that these structures have considerable potential for enhancing non-linear phenomena. Furthermore the non-linear medium need only be present in the small volume within which the energy is concentrated, opening the possibility of using small quantities of expensive material, and reducing any requirements of mechanical integrity that a larger structure would impose.

In passing we draw an analogy with surface enhanced Raman scattering (SERS), observed on rough metallic surfaces, typically silver surfaces. The Raman signal from molecules adsorbed on these surfaces may be enhanced by factors of the order of $10^6$ over that seen on insulating surfaces. The Raman effect is proportional to the second power of the electromagnetic mode density at the surface, and it is known that roughness can enhance the local mode density by factors of up to $10^3 - 10^4$, hence the spectacular Raman enhancement (see [11] for further details and references). A very similar local enhancement takes place in our system and, we expect, can be exploited in an analogous fashion.

In conclusion: we have shown how to design structures made from non-magnetic thin sheets of metal, which respond to microwave radiation as if they had an effective magnetic permeability. A wide range of permeabilties can be achieved by varying the parameters of the structures. Since the active ingredient in the structure, the this metal film, comprises a very small fraction of the volume, typically 1:$10^4$, the structures may be very light, and reinforced with strong insulating material to ensure mechanical strength, without adversely affecting their magnetic properties. It is likely that the structures will be exploited for their ability to concentrate the electromagnetic energy in a very small volume, increasing its density by a huge factor, and greatly enhancing any non-linear effects present.